\documentclass[12pt,a4paper]{conference}
\usepackage{amsfonts}
\newcommand{\la}{\langle}
\newcommand{\ra}{\rangle}
\newcommand{\be}{\begin{equation}}
\newcommand{\eee}{\end{equation}}
\newcommand{\xxxpc}{\%}

\usepackage{fancyhdr}
\usepackage{graphicx,amsmath,amssymb,cite}
\usepackage{multind}
\makeindex{author} \makeindex{subject}

\newcommand{\beq}{\begin{equation}}
\newcommand{\eeq}[1]{\label{#1}\end{equation}}
\newcommand{\eeqn}{\end{equation}}

\newcommand{\beqa}{\begin{eqnarray}}
\newcommand{\eeqa}[1]{\label{#1}\end{eqnarray}}
\newcommand{\eeqan}{\end{eqnarray}}

\let\bar=\overbar

\newcommand{\Dslash}{\not{\hbox{\kern-4pt $D$}}}
\newcommand{\dslash}{\not{\hbox{\kern-2pt $\del$}}}

\newcommand{\msb}{{\bar{\ssstyle M \kern -1pt S}}}

\begin{document}

\Chapter{SCALAR RADIUS OF THE PION AND TWO PHOTONS INTO TWO PIONS. STRONG S-WAVE FINAL STATE INTERACTIONS}
           {Scalar radius of the pion}{J. A. Oller \it{et al.}}
\vspace{-6 cm}\includegraphics[width=6 cm]{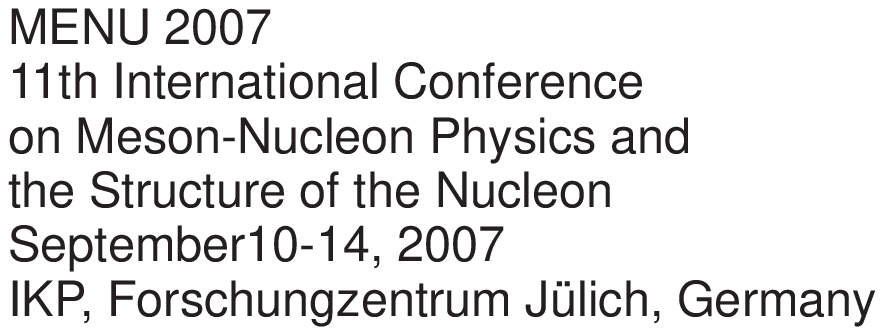}
\vspace{4 cm}

\addcontentsline{toc}{chapter}{{\it N. Author}} \label{authorStart}

\begin{raggedright}

{\it J. A. Oller, L. Roca}\index{author}{J. A. Oller}\\
Departamento de F\'{\i}sica\\
Universidad de Murcia\\
Spain, E-30071
\bigskip\bigskip

{\it C. Schat}\index{author}{C. Schat}\\
Departamento de F\'{\i}sica, FCEyN,\\
Universidad de Buenos Aires\\
Ciudad Universitaria, Pab.1,\\
 (1428) Buenos Aires, Argentina.\\
\bigskip\bigskip

\end{raggedright}

\begin{center}
\textbf{Abstract}
\end{center}
The quadratic pion scalar radius, $\la r^2\ra^\pi_s$,  plays an 
important role for present precise 
determinations of $\pi\pi$ scattering. The solution of the Muskhelishvili-Omn\`es equations for 
 the non-strange null isospin $(I)$ pion scalar form factor determines that
  $\la r^2\ra^\pi_s=0.61\pm 0.04$~fm$^2$. 
 However, by using an Omn\`es representation of this form factor, Yndur\'ain recently obtains 
 $\la r^2\ra^\pi_s=0.75\pm 0.07$~fm$^2$. A large discrepancy between both values, 
  given the precision, then results. We show that Yndur\'ain's method 
  is indeed compatible with the determinations from the Muskhelishvili-Omn\`es equations 
 once a zero in the scalar form factor for some S-wave $I=0$ $T-$matrices is considered. %
   Once this is accounted for, the resulting value is $\la r^2\ra_s^\pi=0.63\pm 0.05$~fm$^2$.
   
  On the other hand, we perform a theoretical study of the reaction 
   $\gamma\gamma\to \pi^0\pi^0$ based on dispersion relations. 
  The  large  
source of uncertainty for $\sqrt{s}\gtrsim 0.5$~GeV, due to variations in the
 phase used in the Omn\`es function above 
the $K\bar{K}$ threshold, 
is removed by taking  one more subtraction in the dispersion relation. 
 This allows us to make sharper predictions for the 
cross section  so that one could use this reaction to 
distinguish between different low energy $\pi\pi$ parameterizations, once independent experiments are 
available.    We also study the role
played by the $\sigma$ or $f_0(600)$ meson in this reaction and determine its width to
two photons.

\section{Introduction}

Here we summarize the two papers \cite{or,ors} that mainly handle with the 
strong influence of the  $I=0$ S-wave  meson-meson final state
interactions. We concentrate here on the non-strange $I=0$ scalar form factor of the pion \cite{or} and 
$\gamma\gamma\to\pi^0\pi^0$  \cite{ors}. Both processes can be formulated in a way
 that has in common the same basic function in order to take care of the strong final state
interactions in the $I=0$ S-wave. This function has been recently the origin of large 
uncertainties  in its
 implementation in the literature, both for the scalar form factor of the pion 
 \cite{y04,y05,y06} and for  $\gamma\gamma\to\pi^0\pi^0$ \cite{penprl}.
 
 The scalar form factor of the pion, $\Gamma_\pi(t)$, corresponds to the matrix element
\be
\Gamma_\pi(t)=\int d^4 x \,e^{-i(q'-q)x}\la \pi(q')|\left(m_u \bar{u}(x)u(x)+ m_d \bar{d}(x)d(x)\right)
| \pi(q)\ra~,~~t=(q'-q)^2~.
\label{ffdef}
\eee

Performing a Taylor expansion around $t=0$, 
\be
\Gamma_\pi(t)=\Gamma_\pi(0)\left\{1+\frac{1}{6}t\la r^2\ra_s^\pi+{\cal O}(t^2)\right\}~,
\eee 
where $\la r^2\ra_s^\pi$ is the quadratic scalar radius of the pion. 
The quantity $\la r^2\ra_s^\pi$ contributes around 10$\xxxpc$  to the values of the 
S-wave $\pi\pi$ scattering lengths $a_0^0=0.220\pm 0.005$ $M_\pi^{-1}$ 
and $a_0^2=-0.0444\pm 0.0010$ $M_\pi^{-1}$, as determined in Ref.~\cite{pipiscat}  
by solving the  Roy equations with constraints from two loop Chiral Perturbation Theory (CHPT). 
If one takes into account that one has 
a  precision of 2.2$\xxxpc$ in the scattering lengths, a 10$\xxxpc$ 
of contribution from $\la r^2\ra_s^\pi$ is a large one.  Related to that, $\la r^2\ra_s^\pi$ 
is also important in $SU(2)\times SU(2)$ CHPT since it gives the 
low energy constant $\bar{\ell}_4$ that controls 
the departure of $F_\pi$ from its value in the chiral limit \cite{gl83,cd04} at next-to-leading order.

Based on one loop $\chi PT$, Gasser and Leutwyler \cite{gl83} obtained $\la r^2\ra_s^\pi=0.55\pm
0.15$~fm$^2$. This calculation was improved later on by the same authors together with 
Donoghue \cite{dgl90}, who 
solved the corresponding Muskhelishvili-Omn\`es equations with the coupled 
channels of $\pi\pi$ and $K\bar{K}$.  The 
update of this calculation, performed in Ref.~\cite{pipiscat}, gives $\la r^2\ra_s^\pi=0.61\pm 0.04$~fm$^2$. 
 Moussallam \cite{m00} employs the same approach and obtains 
values in agreement with the previous result. One should notice that solutions of the
 Muskhelishvili-Omn\`es equations for the scalar form
factor rely on non-measured $T-$matrix elements 
or on assumptions about which are the
channels that matter. Other independent approaches are then most welcome. 
 In this respect we quote the works 
\cite{gu91,uni,chptp6}, and
 Yndur\'ain's ones \cite{y04,y05,y06}. 
These latter works have challenged the previous value for $\la r^2\ra_s^\pi$,
 shifting it to the larger $\la r^2\ra_s^\pi=0.75\pm 0.07$~fm$^2$. If this is translated 
to  the scattering lengths above, employing  an equation of 
Ref.~\cite{pipiscat}, it implies a shift of $+0.006$ $M_\pi^{-1}$ for $a_0^0$ and
$-0.001$ $M_\pi^{-1}$ in $a_0^2$. Thus, one is referring to a shift of 
slightly more than  one sigma. Refs.~\cite{y04,y05} emphasize that one should have a precise knowledge
of the $I=0$  S-wave phase shits, $\delta_0(s)$, for $s\geq 4 M_K^2$~GeV$^2$, $M_K$ is the kaon mass, to disentangle
which of the values, either that of Ref.~\cite{pipiscat} or \cite{y04}, is the right one. However, this point
is based on an unstable behaviour of the solution of Ref.~\cite{y04} with respect to the value of
$\delta_0(4M_K^2)$. Once this instability is cured, as shown below, the resulting $\la r^2\ra_s^\pi$ only 
depends weakly on  $\delta_0(s)$, $s\geq 4M_K^2$, and is compatible with the value of
Ref.~\cite{pipiscat}.

 Regarding the reaction $\gamma\gamma\to\pi^0\pi^0$ one has to emphasize that due to the absence of 
 the Born term (as the $\pi^0$ is neutral), this reaction is specially sensitive to final state interactions. 
For energies  below 0.6~GeV or so, only the S-waves matter, which have $I=0$ or 2.
  It is in this point where both the study of
 this reaction and the scalar form factor match. 
Recently, Ref.~\cite{penprl} updated the dispersive approach of Ref.~\cite{morgan} to calculate 
$\sigma(\gamma\gamma\to\pi^0\pi^0)$. Here one finds a large uncertainty in the 
results for $\sqrt{s}\geq 0.5$~GeV that at around 0.6~GeV is already almost $200\xxxpc$. Again, this is 
due to the lack of a precise 
 knowledge of the phase of the $\gamma\gamma\to\pi\pi$  $I=0$ S-wave amplitude 
above $4m_K^2$. 

We showed in Refs.~\cite{or,ors} that one can improve largely this situation by employing an appropriate Omn\`es
function in the $I=0$ S-wave. The key point is that this function should be continuous under  changes in the
phase functions used above 1~GeV, a point overlooked in the previous studies.

\section{The scalar form factor}

Ref.\cite{y04} makes use of an Omn\`es representation for the pion scalar form factor,
\be
\Gamma_\pi(t)=P(t)\exp\left[\frac{t}{\pi}\int_{4M_\pi^2}^\infty
ds'\frac{\phi_0(s')}{s'(s'-t-i\epsilon)}\right]~.
\label{omnes}
\eee
Here, $P(t)$ is a polynomial in $t$ normalized such that $P(0)=\Gamma_\pi(0)$ and 
 whose zeroes are those of $\Gamma_\pi(t)$. On the other hand, $\phi_0(t)$ is the continuous phase of 
$\Gamma_\pi(t)/P(t)$. Then Refs.~\cite{y04,y05} make use of asymptotic QCD which predicts that the scalar form
factors should go as $-1/t$ times a positive smooth factor for  $t\to +\infty$,
 so that the phase of the form factor should
tend to $+\pi$ in the same limit. At this point, Refs.~\cite{y04,y05} make an assumption that  is not always
necessarily fulfilled. Namely, to identify $\phi_0(t)$ with the phase of $\Gamma_\pi(t)$, that we denote in the
following as $\rho(t)$. If this identification is done, as in Refs.~\cite{y04,y05},
 it follows that $P(t)$ must be a constant, $\Gamma_\pi(0)$, because the behaviour for
$t\to +\infty$ that follows from Eq.~(\ref{omnes}) is 
\be
\Gamma_\pi(t)\to (-1)^{-\phi(\infty)/\pi} t^n t^{-\phi(\infty)/\pi} \Gamma_\pi(0)~,
\eee 
with $n$ the degree of $P(t)$. 
As QCD implies in this assumption 
that $\phi(\infty)/\pi=1$, then $n=0$ and hence $P(t)=\Gamma_\pi(0)$, just a constant.
 One must be aware that in Eq.~(\ref{omnes}) $\phi_0(t)$ is the phase of $\Gamma_\pi(t)/P(t)$.
Notice that the phase of $\Gamma_\pi(t)$ is not continuous when crossing a zero located at 
$t_1\in \mathbb{R}$, as there is a flip in the sign when passing through. However, the phase of $\Gamma_\pi(t)/P(t)$ 
{\it  is} continuous, since the zero is removed. This is the phase one should use in the Omn\`es representation,
Eq.~(\ref{omnes}), because it results from a dispersion relation of $\log \Gamma_\pi(t)/P(t)$, and then $\phi(t)$ must be
continuous (but not necessarily $\rho(t)$).

As stated, Ref.~\cite{y04} took 
\be
\Gamma_\pi(t)=\Gamma_\pi(0)\exp\left[\frac{t}{\pi}\int_{4M_\pi^2}^\infty ds'\frac{\rho(s')}{s'(s'-t-i\epsilon)}\right]~.
\label{ffy04}
\eee
So that the scalar form factor is given by,
\be
\la r^2\ra^\pi_s=\frac{6}{\pi}\int_{4 M_\pi^2}^{+\infty} \frac{\rho(s)}{s^2}ds~.
\label{r2y}
\eee
The phase $\rho(s)$ is fixed in Refs.~\cite{y04,y05} by invoking Watson's final state theorem. For $s<s_K$, $s_K=4
M_K^2$, it implies that $\rho(s)=\delta_0(s)$, where neglecting inelasticity due to multipion states, an experimental fact. 
For $1.42>\sqrt{s}\gtrsim 1.1$~GeV, Ref.~\cite{y04} stressed the interesting fact that experimentally the inelasticity turns 
out  to 
be small and hence Watson's final state theorem can be applied approximately again. In the 
 narrow region between $2M_K$ and 1.1~GeV  inelasticity cannot be neglected but Ref.~\cite{y04} argues that, as it is 
 so narrow, its contribution to Eq.~(\ref{r2y}) is small anyhow and, furthermore, 
that the elasticity parameter $\eta$ is not so small, so that one could still apply Watson's final
 state theorem with corrections.  Thus, for $s_K<s<2$~GeV$^2$, Ref.~\cite{y04} identifies again $\rho(s)\simeq
 \delta_0(s)$. Finally, for $s>s_0=2$~GeV$^2$ Ref.~\cite{y04} takes a linear extrapolation from $\delta_0(s_0)$ to
 $\pi$. One should here criticize that it is still a long way to run from values of $\delta_0(s_0)\lesssim 2\pi$ up to
 $\pi$ at $s\to +\infty$. With all these ingredients, and some error estimates, the value $\la r^2\ra^\pi_s=0.75\pm0.07$~fm$^2$
  results \cite{y04,y05}.
 
 As discussed above in the lines of Ref.~\cite{or}, the steps performed in Ref.~\cite{y04} are not always compatible. In
 Ref.~\cite{or} we took as granted the assumption that Watson's final state theorem
  can be approximately applied for $1.5 \hbox{~GeV}
>\sqrt{s}>2 M_K$. Our assumption is in agreement with any explicit calculation of the pion non-strange $I=0$ scalar form factor
\cite{dgl90,pipiscat,m00,uni} and it is the proper generalized version of the assumption of Refs.~\cite{y04,y05} of identifying $\rho(s)\simeq
\delta_0(s)$. Now, Watson's final state theorem implies that $\phi(s)=\varphi(s)$ (modulo $\pi$),
 with $\varphi(s)$ the phase of the $I=0$ S-wave $\pi\pi$ amplitude, $t_{\pi\pi}=(\eta e^{2 i\delta_0}-1)/2i$. It occurs, as
 stressed in Refs.~\cite{colla,y05}, that $\varphi(s)$ can be either $\sim \delta_0(s)$ or $\sim \delta_0(s)-\pi$ depending on
 whether $\delta_0(s_K)>\pi$ or $<\pi$, respectively, for $s_K<s<2$~GeV$^2$. The latter case corresponds to the calculation in
 Ref.~\cite{pipiscat}, while the former is the preferred one in Ref.~\cite{y05} and arguments are put
 forward for this preference in this reference.
 
 Let us evolve continuously from one situation ($\delta_0(s_K)<\pi$) to the other ($\delta_0(s_K)>\pi$). In the first
 case $\varphi(s)$ has an abrupt drop for $s>s_K$ simply because then $\eta<1$ and while the real part of 
 $t_{\pi\pi}$ rapidly  changes sign, its imaginary part is positive ($>0$). The rapid movement in the real part 
 is due to the swift one in $\delta_0(s)$ in the $K\bar{K}$ threshold due
 to the $f_0(980)$ resonance. As a result for $s\lesssim s_K$, $\varphi(s)=\delta_0(s)\simeq \pi$ and for $s\gtrsim s_K$
 then $\varphi(s)<\pi/2$. This rapid movement gives rise to a rapid drop in the Omn\'es function, 
 Eq.~(\ref{ffy04}), so that the modulus of the form factor has a deep minimum around $s_K$. Here, one is using Watson's final
 state theorem with $\phi_0(s)=\varphi(s)$ and the form factor of Ref.~\cite{dgl90} is reproduced. Notice as well that 
 in this case the function $ \phi(s)$ approaches $\pi$ from below for asymptotic 
 $s$ and then $P(t)=\Gamma_0(0)$ in Eq.~(\ref{omnes}). Now, we consider the limit $\delta_0(s)\to\pi^-$ for $s\to s_K^-$.
 The superscript $-$($+$) indicates that the limit is approached from below(above). 
 In the limit, the change in sign in 
 the real part of $t_{\pi\pi}$ occurs precisely at $s_K$, so that for $s=s_K^-$,
  $\varphi(s)=\pi$ and for $s=s_K^+$ then 
 $\varphi(s)<\pi/2$ (indeed it can be shown from unitarity that must be 0). As a result one has a drop by $-\pi$ in
 $\varphi(s)$ which gives rise to a zero in the Omn\`es representation of the scalar form factor. Thus, the deep has
 evolved to a zero when  $\delta_0(s_K)\to \pi^-$. Because of this zero the proper Omn\`es representation now involves 
  a $P(t)=\Gamma_\pi(0)(1-t/s_K)$ and $\phi(s)$ is no longer $\varphi(s)$ but $\simeq \varphi(s)+\pi\simeq \delta_0(s)$ 
 for $2.25 \hbox{~GeV}^2>s>s_K$. This follows simply because $\phi(s)$ is continuous. Thus, we go into a new realm where
 $\phi(s)\simeq \delta_0(s)$ and the degree of $P(t)$ is 1, so that $\Gamma_\pi(t)$ has a zero at the point $s_1$ where
 $\delta_0(s_1)=\pi$ and $s_1<s_K$. Note that only at $s_1$ the imaginary part of $\Gamma_\pi(t)$ is zero and this fixes 
 the position of the zero \cite{or}. 
 We should emphasize here that if one uses Eq.~(\ref{ffy04}) with  $\phi(s)\simeq 
 \delta_0(s)$, as in Refs.~\cite{y04,y05}, then in the limit $\delta_0(s)\to \pi^+$ for $s\to s_K^+$ the Omn\'es
 representation would give rise to $|\Gamma_\pi(s_K)|=\infty$, while in the previously discussed
 limit of $\delta_0(s)\to \pi^-$ for $s\to s_K^-$ one has $|\Gamma_\pi(s_K)|=0$. 
 This discontinuity was corrected in Ref.~\cite{or}
  and it is the benchmark for a jump by one unit in
 the degree of $P(t)$, a discrete function, in Eq.~(\ref{omnes}).

Hence for $\delta_0(s_K)\geq \pi$ one has to use
\be
\Gamma_\pi(t)=\Gamma_\pi(0)\left(1-\frac{t}{s_K}\right)  
\exp\left[\frac{t}{\pi}\int_{4M_\pi^2}^\infty ds'\frac{\phi(s')}{s'(s'-t-i\epsilon)}\right]~,
\eee
with $\phi(s)\simeq \delta_0(s)$ for $s<2.25$~GeV$^2$. The uncertainties in this approximation for
$s>s_K$ are discussed in Ref.~\cite{or} and included in the final error in $\la r^2 \ra_s^\pi$. The estimation is
based in diagonalizing the $I=0$ S-wave S-matrix for $s<2.25$~GeV$^2$, so that two elastic channels can be singled
out \cite{y05}. We also remark that now  $\phi(s)$ for $\delta_0(s_K)\geq \pi$ must tend to $2\pi$
asymptotically so as to match with the asymptotic behaviour of $\Gamma_\pi(t)$ as $-1/t$. In this way we have
now a very soft matching with asymptotic QCD  since for $s$ around 2.25~GeV$^2$, $\delta_0(s)\simeq 2 \pi$. 
 This was not the case in Ref.~\cite{y04,y05}. Notice that
from our work it follows that the precise knowledge of the asymptotic behaviour of the phase of the form factor
is not relevant as $\phi(s)$ can tend either to $2\pi$ ($\delta_0(s_K)>\pi$) or to $\pi$ ($\delta_0(s_K)<\pi$), and the
results are very similar. 

Our final value is 
\be
\la r^2\ra_s^\pi=0.63 \pm 0.05~\hbox{fm}^2.
\eee
The error takes into account different $\pi\pi$ $I=0$ S-wave parameterizations, namely those of
Refs.~\cite{pipiscat} and \cite{py03}, the error in the application of Watson's final state theorem
above 1~GeV  
and up to 1.5~GeV, and the uncertainties in $\phi(s)$ given by asymptotic QCD for $s>2.25$~GeV$^2$.
 This value is compatible with that of Ref.~\cite{pipiscat}, $\la r^2\ra_s^\pi=0.61\pm 0.04$~fm$^2$, and also with
 $\la r^2\ra_s^\pi=0.64\pm 0.06$~fm$^2$ of Ref.~\cite{uni} calculated from  Unitary CHPT.

\section{The $\gamma\gamma\to\pi^0\pi^0$ reaction}

In this section we briefly review Ref.~\cite{ors}. This reference extended the approach of
Refs.~\cite{penprl,morgan} so as to be less sensitive to the phase of the  $I=0$ S-wave
$\gamma\gamma\to\pi\pi$ amplitude above $s_K$. For this phase one has a similar situation to that of the 
scalar form factor of the pion, it can be either $\sim \delta_0(s)$ or $\sim \delta_0(s)-\pi$ 
for $1\lesssim s \lesssim 2.25$~GeV$^2$ \cite{ors,penprl}. In the approach of Ref.~\cite{penprl} this originates  
an uncertainty that raises dramatically with energy above 0.5~GeV, such for $\sqrt{s}\simeq 0.6$~GeV it is
already 200$\xxxpc$.

Let us denote by $F_I(s)$ the S-wave $I=0$ $\gamma\gamma\to\pi\pi$ amplitude. 
The approach of Ref.~\cite{penprl,morgan} is based on isolating the left hand cut contribution of $F_I$
which is denoted by $L_I$. These authors also employ the Omn\`es function
\be
\Omega_I(s)=\exp\left[\frac{s}{\pi}\int_{4 M_\pi^2}^{+\infty}ds'\frac{\phi_I(s')}{s'(s'-s)}\right]~,
\label{omegapen}
\eee
where $\phi_I(s')$ is the phase of $F_I(s)$. For $I=2$ by the application of Watson's
 final state theorem one has that $\phi_2(s)=\delta_2(s)$. For $I=0$ and $s<s_K$, 
 $\phi_0(s)=\delta_0(s)$. In the interval $1.5>\sqrt{s}>1.1$~GeV, $\phi_0=\delta_0$ (modulo $\pi$) because 
 inelasticity is small again, as already remarked. 
 Similarly as in the scalar form factor one can have because of the onset of inelasticity 
 above $2 M_K$ and up to 1.1~GeV, that $\phi_0$ is given either by $\sim \delta_0$ or 
  $\sim \delta_0-\pi$.  
  
Ref.~\cite{penprl} then performed a twice subtracted 
 dispersion relation of the function
$(F_I(s)-L_I(s))/\Omega_I(s)$. An important point to realize is that the previous function
 has no left hand cut and that  $F_I/\Omega_I$ has no right hand cut. Making use of the  Low's
theorem, which implies that  $L_I(s)$ is given by the Born term $B_I(s)$ for $s\to 0$, one
is only left with two subtraction constants to be fixed. One of these
 constants can be fixed by
requiring that the $\gamma\gamma\to\pi^0\pi^0$ S-wave amplitude, $F_N(s)$,
 has an  Adler zero around $M_\pi^2$. The other one was fixed in Ref.~\cite{penprl} by
 requiring that the $\gamma\gamma\to \pi^+\pi^-$ S-wave amplitude, $F_C(s)$, tends to the
 Born term $B_C(s)$ for $s\to 0$ up to ${\cal O}(s^2)$. One has to say that Ref.~\cite{penprl} did
 not include axial vector exchanges which indeed give rise to a  term that vanishes for 
 $s\to 0$ only linearly in $s$. This gives rise to a difference in the cross section of
 around a $30\xxxpc$ at $\sqrt{s}\simeq 0.5$~GeV.
 
 In order to better handle the ambiguities in $\phi_0(s)$ above 1~GeV, Ref.~\cite{ors} only 
 uses  $\Omega_0(s)$ of Eq.~(\ref{omegapen})\footnote{We already know about the lack of
 continuity of $\Omega_0(s)$ when $\delta_\pi(s_K)$ crosses $\pi$ when taking $\phi_0(s)$
 given by $\varphi(s)$ as in the case of the scalar form factor.} 
 for $\phi_0(s)\sim \delta_0(s)-\pi$ for $s>1$~GeV$^2$. For the case $\phi_0(s)\sim
 \delta_0(s)$ above 1~GeV Ref~.\cite{ors} employs 
 \be
\widetilde{\Omega}_0(s)=\left(1-\frac{s}{s_1}\right)
 \exp\left[\frac{s}{\pi}\int_{4 M_\pi^2}^{+\infty}ds'\frac{\phi_I(s')}{s'(s'-s)}\right]~,
 \label{omnesors}\eee
and then a twice dispersion relation of $(F_0-L_0)/\widetilde{\Omega}_0$ is  performed.
It is important to realize, as stressed in Ref.~\cite{ors}, that because 
of the first order polynomial in front of the exponential in Eq.~(\ref{omnesors}), one
indeed has a {\it three} times subtracted dispersion relation for 
$(F_0-L_0)/\Omega_0$. Recall that the latter is the original function used in 
Refs.~\cite{penprl,morgan}.

Because of this extra subtraction  one can reduce dramatically the sensitivity to the
 $\phi_0(s)$ above 1~GeV. The conditions used to fix the at most three subtraction
 constants that appear in our scheme 
  are: i) $F_N(s)\to 0$ for $s\to 0$ with the slope fixed by one loop
CHPT \cite{bijcornet}
 (with an uncertainty of around $15\%$), ii) $F_C(s)\to
B_C(s)+{\cal O}(s)$ with the rest fixed by one loop CHPT (with the same $15\xxxpc$ of estimated uncertainty).
 The third condition is an upper bound to the value of the resulting cross section in the $f_0(980)$ region 
 so that it is smaller
than 200 nb. Notice that its experimental value is smaller than 40 nb and, hence, we take here a 
very conservative uncertainty.

\begin{figure}[Ht]
\begin{center}
\includegraphics[height=7 cm, width=12.5 cm]{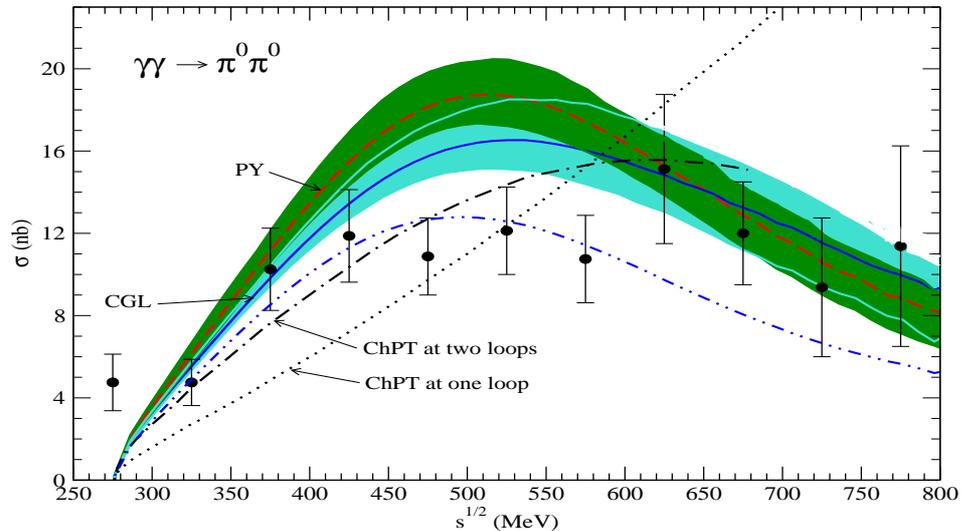}
\caption{\protect \small
Final results for the $\gamma\gamma\to\pi^0\pi^0$ cross
section. Experimental data are from the Crystal Ball Coll. \cite{ball},
 scaled by $1/0.8$, as   $|\cos\theta|<0.8$ is measured and S-wave dominates. 
 The lighter band corresponds to Ref.~\cite{pipiscat} while the darker one 
 to Ref.~\cite{py03}. 
The dot-dot-dashed line results after removing the axial vector exchange contributions,
 as in Ref.~\cite{penprl} with $\phi_0(s)\sim \delta_0(s)-\pi$ for $s>1$~GeV$^2$.
 The band along each line represents the theoretical uncertainty. The dotted line 
 is the one loop $\chi$PT result \cite{bijcornet} 
 and the dot-dashed one the two loop calculation  
 \cite{sainioetal}. } \label{figgamm}
\end{center}
\end{figure}

We show in Fig.~\ref{figgamm} our results together with  the experimental
points from Ref.~\cite{ball}. The darker band corresponds to employ Ref.~\cite{py03}
for $\delta_0(s)$ below 1~GeV and the lighter one to use Ref.~\cite{pipiscat}. One sees 
that now with more precise data one should be able
to distinguish between different low energy $\delta_0(s)$ parameterizations as the theoretical uncertainty is much reduced. 
 The widths of the bands correspond to the 
uncertainties related to the $\delta_0(s)$ and $\delta_2(s)$ parameterizations used, 
those in fixing the three
subtraction constants and in employing Watson's final state theorem for $s>1 $~GeV$^2$, 
 and it also includes  the uncertainty in the asymptotic
$\phi_I(s)$ employed. In the figure we also
show with the dotted line the one loop CHPT result \cite{bijcornet} and with the dash-dotted line 
the two loop one
 \cite{sainioetal}. There is a clear improvement when going from one to two loops in  CHPT,
 though to have a perfect agreement with our results some higher order corrections are still
 needed. Finally, the dash-double-dotted line corresponds to the result of Ref.~\cite{penprl}
  with $\phi_0(s)\sim \delta_0(s)-\pi$ for $s>1$~GeV$^2$. Let us recall that Ref.~\cite{penprl}
 does not include axial vector exchanges. Were they included, the
results of this reference would fall inside the bands shown by our results. 
 
 By analytical continuation on the complex plane one can determine the coupling of the
 $\sigma$ to $\gamma\gamma$, $g_{\sigma\gamma\gamma}$,
  and calculate the width to $\gamma\gamma$ of this resonance
 \cite{ors}. We then obtain for the ratio of couplings 
 $\left|\frac{g_{\sigma\gamma\gamma}}{g_{\sigma\pi\pi}}\right|=(2.1\pm 0.2)\times 10^{-3}$~,
 with $g_{\sigma\pi\pi}$ the $\sigma$ coupling to two pions. 
 The result of \cite{penprl} corresponds to this ratio  being $20\%$ bigger at $(2.53\pm 0.09)\times 10^{-3}$.  Half of this
difference is due to the omission of the exchanges of axial vector resonances in \cite{penprl}, and
the other half comes from improvements delivered by our extra subtraction and our slightly
 different inputs. As a result, using the same value for $|g_{\sigma \pi\pi}|$ as in 
 \cite{penprl}, our resulting value for $\Gamma(\sigma\to\gamma\gamma)$ would be around 
 a $40\%$ smaller  than that in \cite{penprl}.  
 Taking into account different choices of $|g_{\sigma\pi\pi}|$ we end with $\Gamma(\sigma\to\gamma\gamma)$ 
in the interval $1.8-3$~KeV.
   
\section{Conclusions}

We have shown that both Yndur\'ain's method \cite{y04} and the solution of the
Muskhelishvili-Omn\'es equations \cite{dgl90,pipiscat} provide compatible
results for the quadratic scalar radius of the pion. The origin of the
discrepancy between Refs.~\cite{y04} and \cite{pipiscat} was due to overlooking a
zero in the scalar form factor in the former reference. We finally obtain
\cite{or} $\la r^2\ra_s^\pi=0.63 \pm 0.05~\hbox{fm}^2$ and $\bar{\ell}_4=4.5\pm
0.3$. These numbers are in good agreement with $\la r^2\ra_s^\pi=0.61\pm 0.04$~fm$^2$
 and $\bar{\ell}_4=4.4\pm 0.2$ of Ref.~\cite{pipiscat}.
 
 We have also studied the $\gamma\gamma\to \pi^0\pi^0$ reaction for energies
 $\sqrt{s}\lesssim 0.7$~GeV, where S-waves dominate. We have extended the original approach 
 of Ref.~\cite{morgan,penprl} by performing a three times subtracted dispersion relation \cite{ors}, 
 instead of the
 twice subtracted originally employed. The sensitivity 
of the results with respect to the phase of the $I=0$   $\gamma\gamma\to\pi\pi$ 
 S-wave  above 
$4 M_K^2$ is then largely reduced. A key point is to properly handle the contribution of the
$f_0(980)$ resonance, at least at the level of the order of magnitude.
 Importantly, one can then use
this reaction to distinguish between different low energy $\pi\pi$ parameterizations once new data on
$\sigma(\gamma\gamma\to \pi^0\pi^0)$ are available. The $\Gamma(\sigma\to \gamma\gamma)$ width is estimated
 in the range $1.8-3$~KeV \cite{ors}.

\section*{Acknowledgments}

 This work has been supported in part by the MEC (Spain) and FEDER (EC) Grants
  FPA2004-03470 and Fis2006-03438,  the 
  Fundaci\'on  S\'eneca (Murcia) grant Ref. 02975/PI/05, the European Commission
(EC) RTN Network EURIDICE  Contract No. HPRN-CT2002-00311 and the HadronPhysics I3
Project (EC)  Contract No RII3-CT-2004-506078. 



\begin{thebibliography}{0}    

\bibitem{or}  J.~A.~Oller and L.~Roca, {\it Phys.\ Lett.}\   {\bf B651}, 139 (2007).

\bibitem{ors}   J.~A.~Oller, L.~Roca and C.~Schat,  {\it arXiv:0708.1659 [hep-ph].} To appear 
in Phys. Lett. {\bf B}.

\bibitem{y04}F. J. Yndur\'ain, {\it Phys. Lett.} {\bf B578}, 99 (2004); (E)-$ibid$ {\bf B586}, 439 (2004).
 
\bibitem{y05}F. J. Yndur\'ain, {\it Phys. Lett.} {\bf B612}, 245 (2005).

\bibitem{y06}F. J. Yndur\'ain, {\it arXiv:hep-ph/0510317.}

\bibitem{penprl} M.~R.~Pennington, {\it Phys.\ Rev.\ Lett.}  {\bf 97}, 011601 (2006). 


\bibitem{pipiscat} G. Colangelo, J. Gasser and H. Leutwyler, {\it Nucl. Phys.} {\bf B603}, 125 (2001).  

\bibitem{gl83}J. Gasser and H. Leutwyler, {\it Phys. Lett.} {\bf  B125}, 325 (1983).

\bibitem{cd04}G. Colangelo and S. D\"ur, {\it Eur. Phys. J.} {\bf C33}, 543 (2004). 

\bibitem{dgl90} J. F. Donoghue, J. Gasser and H. Leutwyler, {\it Nucl. Phys.} {\bf B343}, 341 (1990). 

\bibitem{m00}B. Moussallam, {\it Eur. Phys. J.} {\bf C14}, 111 (2000). 

\bibitem{gu91}J. Gasser and U.-G. Mei{\ss}ner, {\it Nucl. Phys.} {\bf B357}, 90 (1991). 
 
\bibitem{uni}U.~G.~Mei{\ss}ner and J.~A.~Oller, {\it Nucl.\ Phys.}  {\bf A679}, 671 (2001).   

\bibitem{chptp6} J. Bijnens, G. Colangelo and P. Talavera, {\it JHEP} {\bf 9805}, 014 (1998). 

\bibitem{morgan}D.~Morgan and M.~R.~Pennington,
  {\it Phys.\ Lett.}  {\bf B272}, 134 (1991); {\it Z.\ Phys.}   {\bf C37}, 431 (1988)
  [Erratum-ibid.\  C {\bf 39}, 590 (1988)].

\bibitem{colla} B.~Ananthanarayan, I.~Caprini, G.~Colangelo, J.~Gasser and H.~Leutwyler,
   {\it Phys.\ Lett.}   {\bf B602}, 218 (2004).

\bibitem{py03}J. R. Pel\'aez and F. J. Yndur\'ain, Phys. Rev. {\bf D68}, 074005 (2003); 
$ibid$ {\bf D71}, 074016 (2005). 

\bibitem{bijcornet}J. Bijnens and F. Cornet, {\it Nucl. Phys.} {\bf B296}, 557 (1988).
 J. F. Donoghue, B. R. Holstein and Y. C. Lin, {\it Phys. Rev.} {\bf D37}, 2423 (1988).

\bibitem{ball} H.~Marsiske {\it et al.} [Crystal Ball Collab.],
  Phys.\ Rev.\   {\bf D41},  3324 (1990).
  
\bibitem{sainioetal}S. Bellucci, J. Gasser and M. E. Sainio, Nucl. Phys. {\bf B423}, 80
(1994); J. Gasser, M. A. Ivanov and M. E. Sainio, Nucl. Phys. {\bf B728}, 31 (2005).



\end{thebibliography}
\end{document}